\documentclass[12pt]{article}
\usepackage{graphicx}
\usepackage{amssymb,amsmath}

\def\warwick{Department of Physics, The University of Warwick, \\
Gibbet Hill Road, Coventry, CV4 7AL, UK}
\def\binp{Budker Institute of Nuclear Physics\\
Lavrentieva 11, Novosibirsk, 630090, Russia}

\def\Title#1{\begin{center} {\Large #1 } \end{center}}
\def\Author#1{\begin{center}{ \sc #1} \end{center}}
\def\Address#1{\begin{center}{ \it #1} \end{center}}

\newenvironment{Abstract}{\begin{quotation}  }{\end{quotation}}
\newenvironment{Presented}{\begin{quotation} \begin{center} 
             PRESENTED AT\end{center}\bigskip 
      \begin{center}\begin{large}}{\end{large}\end{center} \end{quotation}}

\renewcommand{\deg}{{}^{\circ}}
\newcommand{\ee}{$e^+e^-$}

\begin{document}
\begin{titlepage}

\vfill
\Title{Ultimate sensitivity on $\gamma/\phi_3$ from $B\to DK$}
\vfill
\Author{Anton Poluektov}
\Address{\warwick\\ \binp}
\vfill
\begin{Abstract}
Measurement of the CKM phase $\gamma$ in $B\to DK$ decays can be potentially 
performed with high precision due to low theoretical uncertainties. 
However, the precision measurement requires very large experimental samples 
of $B$ decays. This report covers prospects for $\gamma$ measurement
at the future $e^+e^-$ facilities and upgraded LHCb detector. 
\end{Abstract}
\vfill
\begin{Presented}
6th International Workshop on the CKM Unitarity Triangle\\
University of Warwick, UK, 6--10 September 2010
\end{Presented}
\vfill
\end{titlepage}
\def\thefootnote{\fnsymbol{footnote}}
\setcounter{footnote}{0}

\section{Introduction}

Measurement of the CKM phase $\gamma$ (also known as $\phi_3$) 
in $B\to DK$ decays can be potentially 
performed with high precision due to low theoretical uncertainties. 
Since no loop-level diagrams are involved, this measurement provides 
a Standard Model reference for other determinations of CKM parameters. 
However, the precision measurement requires very large samples 
of $B$ decays. Presented here are simple estimates of $\gamma$ sensitivity 
in the next-generation experiments. 

Two projects of the future \ee\ facilities are now approved: SuperB project
in Italy~\cite{superb_cdr} and upgraded
KEKB collider (SuperKEKB) with the Belle II detector in KEK laboratory 
in Tsukuba, Japan~\cite{belle2_tdr}. SuperB has design luminosity of 
$10^{36}$ cm$^{-2}$s$^{-1}$, 
and aims to accumulate 75 ab$^{-1}$ during its operation. SuperKEKB
has similar design goals ($8\times 10^{35}$ cm$^{-2}$s$^{-1}$ and 50 ab$^{-1}$, 
respectively). Both projects plan to start data taking around
2015 and continue until 2020. 

LHCb experiment~\cite{lhcb} is a facility that will make use 
of large $B$ production cross section in proton collisions at LHC 
to study $B$ decays. After the first phase of experiments in 2009--2016 
when the sample of 6--7 fb$^{-1}$ will be accumulated, there are plans to  
upgrade LHCb to be able to run with the increased luminosity of 
$2\times 10^{33}$ cm$^{-1}$s$^{-1}$. The expected 
data sample to be collected by LHCb by 2020 is 50--100 fb$^{-1}$. 

The following estimates assume luminosity integrals 
of 50 ab$^{-1}$ for \ee\ machines operated at $\Upsilon(4S)$
(we will refer to both projects as SuperB), 
and 50 fb$^{-1}$ for upgraded LHCb at the centre-of-mass energy of 14 TeV. 
These integrals roughly correspond to $50\times 10^9$ produced $B\overline{B}$
pairs at the SuperB and $20\times 10^{12}$ $B\overline{B}$ pairs at LHCb. 
Expected numbers of events in the benchmark modes for $\gamma$ measurement
are shown in Table~\ref{num_ev}. The numbers have been
obtained by rescaling the signal yields at B-factories and from MC 
simulation studies of LHCb~\cite{roadmap}. First results from LHCb show 
that the yields are roughly consistent with the MC expectation. 

\vspace{-\baselineskip}
\begin{table}[hb!]
\caption{Numbers of events in benchmark modes for $\gamma$ measurement}
\label{num_ev}
\centering
\begin{tabular}{|l|ll|}
  \hline
  Mode                    & SuperB (50 ab$^{-1}$) & LHCb (50 fb$^{-1}$) \\
  \hline
  $B\to D(K\pi)K$ allowed & 200K                  & 4M   \\
  $B\to D(K_S\pi\pi)K$    & 100K                  & 300K \\
  $B_s\to D_s(KK\pi)K$    & -                     & 700K \\
  \hline
\end{tabular}
\end{table}
\vspace{-\baselineskip}

\section{Methods involving \boldmath{$D\to hh$}}

At SuperB, conventional GLW method can be applied that involves 
decays of $D$ to $CP$-even ($h^+h^-$) and $CP$-odd ($K^0_S\pi^0$, $K^0_S\omega$)
final states. Four observables are 
\begin{equation}
R_{\pm} = 1+r_B^2\pm 2r_B\cos\delta_B\cos\gamma, \;\;
A_{\pm} = \pm 2r_B\sin\delta_B\sin\gamma/R_{\pm}, 
\end{equation}
where $R_{\pm}$ is the ratio of $CP$ to flavor-specific 
branching ratios, and $A_{\pm}$ is the relative charge asymmetry. 
The observables are not independent: $R_{+}A_{+}=-R_{-}A_{-}$, 
but there is still enough information to extract the unknown parameters
$\gamma$, amplitude ratio $r_B$ and strong phase $\delta_B$. 
In practise, the sensitivity of GLW method
alone is poor and there is a 4-fold ambiguity: 
$(\gamma,\delta_B)\to (\pi-\gamma,\pi-\delta_B)$ and 
$(\gamma,\delta_B)\to (\delta_B, \gamma)$. 

Adding ADS mode that involves doubly Cabibbo-suppressed decay $D^0\to
K^+\pi^-$ helps improve the statistical sensitivity and resolve
ambiguities. The observables are
\begin{equation}
R_{ADS} = r_B^2 + r_D^2 + 2r_Br_D\cos\delta\cos\gamma, \;\;
A_{ADS} = 2r_Br_D\sin\delta\sin\gamma/R_{ADS}, 
\end{equation}
where $\delta$ is a sum of $\delta_B$ and the phase difference $\delta_D$
between $\overline{D}{}^0$ and $D^0\to K^-\pi^+$ amplitudes, 
and $r_D$ is the ratio of these amplitudes. 
The ambiguity $(\gamma,\delta_B)\to (\pi-\gamma,\pi-\delta_B)$, 
however, is not resolved without external constraints on 
$\delta_D$ value. This constraint also improves the statistical precision 
significantly. Measurement of $\delta_D$ has been performed by 
CLEO~\cite{cleo_deltad} ($\delta_D=(22^{+11}_{-12}{}^{+9}_{-11})^{\circ}$), 
but better accuracy is required for a precision $\gamma$
measurement. The use of other $D$ decay modes, such as $K\pi\pi\pi$ or 
$K\pi\pi^0$, gives additional constraints on $r_B$ which also 
improves $\gamma$ precision. 

The application of methods involving two-body $D$ decays at LHCb is 
different, since $CP$-odd final states are not readily available. 
LHCb will therefore use a combination of GLW ($D\to KK, \pi\pi$) and 
ADS ($D\to K\pi$) modes to constrain the free parameters $\gamma$, 
$r_B$, $\delta_B$~\cite{roadmap}. The numbers of events in these modes are: 
\begin{equation}
  \begin{split}
    N(B^{\pm}\to D(K^{\pm}\pi^{\mp})K^{\pm}) = &
    N_{K\pi}[1+r_B^2r_D^2 + 2r_Br_D\cos(\delta_B-\delta_D\pm\gamma)],\\
    N(B^{\pm}\to D(K^{\mp}\pi^{\pm})K^{\pm}) = &
    N_{K\pi}[r_B^2 + r_D^2 + 2r_Br_D\cos(\delta_B+\delta_D\pm\gamma)],\\
    N(B^{\pm}\to D(h^+h^-)K^{\pm}) = &
    N_{hh}[1+r_B^2 + 2r_B\cos(\delta_B\pm\gamma)]. 
  \end{split}
\end{equation}
If the ratio of normalisation factors $N_{K\pi}/N_{hh}$ is fixed 
(using the efficiency ratio obtained from MC or control samples), 
these equations provide enough constraints to obtain unknown 
parameters $r_{B}$, $\delta_{B,D}$ and $\gamma$. 

Expected $\gamma$ sensitivity for $D\to hh$ modes is given in Table~\ref{dhh}. 
The numbers are obtained for $\gamma=70^{\circ}$, $r_B=0.1$, $\delta_B=130^{\circ}$, 
$\delta_D=22^{\circ}$. 
Significant disadvantage of the method involving $D\to hh$ decays is in 
its precision being dependent on the values of the strong phases. 
The precision degrades significantly for the combinations of phases
$\delta_B+\delta_D\pm\gamma$ close to 0 or $\pi$ --- in these cases 
the number of events in ADS mode becomes insensitive to the values of phases, 
and one needs to use additional constraints ({\it e. g.} other 
$D$ decay modes, such as $K\pi\pi\pi$, to constrain $r_B$). 

\begin{table}[t]
  \caption{Estimated sensitivity of SuperB and LHCb to $\gamma$ 
           using GLW and ADS modes}
  \label{dhh}
  \centering
  \begin{tabular}{|l|c|c|}
  \hline
                           & SuperB (50 ab$^{-1}$) & LHCb (50 fb$^{-1}$) \\
  \hline
  $D\to hh$, $D\to K\pi$                           & $5.1\deg$   & $1.4\deg$ \\
  $D\to hh$, $D\to K\pi$, $\sigma(\delta_D)=1\deg$ & $3.9\deg$   & $1.0\deg$ \\
  $D\to hh$, $D\to K\pi$, $D\to K\pi\pi\pi$        & $2.4\deg$   & $0.8\deg$ \\
  \hline
  \end{tabular}
\vspace{-\baselineskip}
\end{table}

Additional modes with neutrals can be used at SuperB: 
$B^{\pm}\to D^*(D\pi^0, D\gamma)K^{\pm}$, $B^{\pm}\to DK^{*\pm}$. Using these
modes can double the $\gamma$ precision and provides a fallback solution 
for the ``unlucky'' phase combinations in $B^{\pm}\to DK^{\pm}$. 

\section{Dalitz plot analyses}

The $\gamma$ measurement using the Dalitz plot analysis of $D\to K^0_Shh$ decay 
provides the best constraint on $\gamma$ with $B$-factories data, but in the
case of LHCb its contribution is suppressed by relatively low 
trigger efficiency for modes involving $K^0_S$ reconstruction. 
The observables in this method are Dalitz plot densities of $D$ decay from 
$B^{\pm}\to DK^{\pm}$: 
\begin{equation}
  p_{\pm}(m_{K^0_S\pi^+}^2, m_{K^0_S\pi^-}^2) = 
   |A_D(m_{K^0_S\pi^+}^2, m_{K^0_S\pi^-}^2) + 
   r_Be^{i(\delta_B\pm \gamma)}A_D(m_{K^0_S\pi^-}^2, m_{K^0_S\pi^+}^2)|^2. 
\end{equation}
The advantage of this method is in its precision being only weakly 
dependent on the values of phases because of significant strong phase 
variation in the amplitude $A_D$. 

The analyses performed so far use $D\to K_S^0\pi^+\pi^-$
and $D\to K_S^0K^+K^-$ amplitudes obtained from the flavor-tagged sample 
$D^{*\pm}\to D^0\pi^{\pm}$ with model assumptions. The associated model 
uncertainty reaches a few degrees, and would dominate the future 
precision measurements. To avoid it, a modification of the Dalitz plot analysis 
method has been proposed, where all the information about the amplitude 
$A_D$ is obtained from experiment. If the Dalitz plot is divided into bins
$i$ (symmetric to the exchange of $\pi$ charge which 
corresponds to $i\leftrightarrow -i$), the number of events in the bin $i$
of the Dalitz plot from $B^{\pm}\to DK^{\pm}$ decays is 
(with arbitrary normalisation)
\begin{equation}
  M^{\pm}_i = K_i + r_B^2K_{-i}+2\sqrt{K_iK_{-i}}(x_{\pm}c_i + y_{\pm}s_i), 
\end{equation}
where $K_i$ is the number of events in the bin $i$ of flavor $D^0$ Dalitz 
plot, and $x_{\pm}+iy_{\pm} = r_Be^{i(\delta_B\pm \gamma)}$. Average cosine $c_i$ and sine $s_i$ of the strong phase 
difference between the bins $i$ and $-i$ can be obtained from the analysis 
of quantum-correlated $\psi(3770)\to D^0\overline{D}{}^0$ decays. 
Using the binning inspired by the measured $D^0$ amplitude 
allows to reach the precision only $10-20\%$ worse than 
in the unbinned approach~\cite{modind}. 

Extrapolation of the sensitivity reached at the B-factories to the 
SuperB and LHCb data samples gives the statistical 
precision of 1--2$^{\circ}$. This clearly means that the analysis should 
be done in a model-independent way. The precision of $c_i$, $s_i$
obtained by CLEO~\cite{cleo_cs} with 800 pb$^{-1}$ translates 
to 2--3$^{\circ}$ error in $\gamma$. 
Therefore one needs a larger (of the order 10 fb$^{-1}$) 
sample of $e^+e^-\to \psi(3770)$ data that 
can be provided by {\it e. g.} BES-III experiment. 
 
Although $D\to K_S\pi\pi$ presently dominates the sensitivity among the 
multibody modes, other three-body (and possibly four-body) 
$D^0$ final states can be used for $\gamma$ extraction using 
Dalitz plot analyses: $K^0_SK^+K^-$ ($c_i$, $s_i$ measurement by CLEO is 
available for this mode~\cite{cleo_cs}) , $K^0_SK\pi$, $\pi^+\pi^-\pi^0$, 
$K^+K^-\pi^+\pi^-$, $K^0_S\pi^+\pi^-\pi^0$ etc. 

\section{Time-integrated \boldmath{$B^0$} analyses}

Self-tagged decays of the neutral $B$ ($B^0\to DK^{*0}$) can be used 
in the same way as the $B^{\pm}\to DK^{\pm}$ decays. The branching ratio of this mode 
is smaller by a factor 4--5, but since both interfering amplitudes are
colour-suppressed, larger CP violation with $r_B\sim 0.3$ is expected, 
thus giving the $\gamma$ sensitivity similar to $B^+\to DK^+$ mode. 
A complication in $B^0\to DK^{*0}$ analyses is in $K^{*0}$ being a wide 
state. An interference with other states in $B^0\to DK\pi$ amplitude 
affects the results of the $\gamma$ measurement. This effect can be corrected
in a model-independent way by introducing an additional free parameter --- 
coherence factor $\kappa<1$ in the interference term. 

This complication turns into advantage if one performs a simultaneous
Dalitz plot analysis of $B^0\to DK\pi$ amplitudes with $D\to K\pi$ and $D\to hh$
final states~\cite{bdkp}. The interference with the flavor-specific 
$D^*_2K$ mode 
resolves discrete ambiguities and provides additional sensitivity 
compared to ADS- and GLW-like techniques. An attempt has been made to make this 
approach model-independent using the double Dalitz plot analysis 
$B^0\to D(K^0_S\pi^+\pi^-)K\pi$~\cite{bdkp_modind}. Feasibility studies 
show the precision of $\gamma$ measurement with upgraded LHCb 
around 1--1.5$^{\circ}$, but this number has a large uncertainty since 
the amplitudes $B^0\to DK\pi$ are not well-measured yet. 

\section{Time-dependent analyses}

In the time-dependent analyses, $\gamma$ is obtained 
from the interference of $B^0\to D^{(*)-}\pi^+$ and 
$\overline{B}{}^0\to D^{(*)-}\pi^+$ amplitudes through $B$ mixing. 
Although the $CP$ violation in now well-established in this mode, precise 
determination of $\gamma$ is difficult due to smallness of the amplitude 
containing $V_{ub}$. In $B_s\to D_s^{\pm}K^{\mp}$
decays accessible at LHCb both interfering amplitudes are of the same order, 
which makes them competitive to the time-integrated $B\to DK$ 
measurements~\cite{roadmap}. 
Low cross section of $B_s$ production in $e^+e^-$ collisions does 
not allow SuperB to compete here. 

Measurements of flavor-tagged $B_s\to D_sK$ decay rate as a function of 
time are sensitive to $\gamma+\phi_s$. The $CP$-violating phase $\phi_s$
will be measured precisely in $B_s\to J/\psi\phi$ analysis; no other 
external inputs are needed for $\gamma$ measurement. MC studies give 
the sensitivity of $10^{\circ}$ with 2 fb$^{-1}$ at LHCb, which scales 
to 1.5--2$^{\circ}$ with upgraded LHCb data. It still needs to be understood, 
however, if there are systematic effects that can limit this precision. 

\section{Influence of charm mixing}

Mixing in charm sector can affect the precision $\gamma$ measurements 
that use neutral $D$ mesons in the final state. The magnitude of the 
effect depends on how the charm data are used in $\gamma$ measurement. 
All ADS and Dalitz $B\to DK$ analyses performed so far ignore $D$ mixing 
in both charm and $B$ parameters. In that case, the mixing correction 
is of the second order in 
$x_D, y_D$ and thus can be safely ignored even for one-degree $\gamma$ 
precision~\cite{gamma_dmix}. Model-independent Dalitz plot analysis is however a 
special case: $c_i$ and $s_i$ parameters extracted from quantum-correlated 
$\psi(3770)\to D^0\overline{D}{}^0$ decays where neutral $D$ mesons are 
produced in $C=-1$ state appear to be unaffected by mixing~\cite{modind_mix}. 
Using these values in $B\to DK$ analysis results in linear mixing effects in 
$\gamma$ that are, however, additionally suppressed by $r_B$ and 
other small factors. The resulting $\gamma$ bias is estimated to be 
of the order $0.2^{\circ}$ and can be corrected once the $D$ mixing 
parameters will be measured. 

\section{Conclusion}

Various independent methods allow measurements of CKM phase $\gamma$
with the precision of 1--3$^{\circ}$ at the future $e^+e^-$ and hadron 
facilities. While the upgraded LHCb sensitivity potentially looks more 
promising than that of SuperB, it can be affected by higher backgrounds 
and degenerate combinations of parameters. In addition, both facilities
have their own auxiliary modes: $D^*$ states (SuperB), $B_s$ and $\Lambda^0_b$
decays (LHCb). 

Having a large $10$ fb$^{-1}$ sample of $\psi(3770)\to D\overline{D}$
decays is desirable for the model-independent extraction of hadronic 
parameters in $D$ decays. This can be achieved by taking a 
significant fraction of BES-III sample at open charm threshold, by 
building a high-luminosity charm-tau factory, or by running SuperB at 
low energy. 

I am grateful to Tim Gershon and Robert Fleischer for valuable comments. 
This work is supported by the Science and Technology Facilities Council 
(UK).

\end{document}